\documentclass[sigplan,screen]{acmart}

\usepackage{svg}

\usepackage{amsmath}
\DeclareMathOperator*{\argmax}{arg\,max}

\usepackage{graphicx}
\usepackage{textcomp}
\usepackage{xcolor}

\usepackage{algorithm}
\usepackage{algpseudocode}

\usepackage{tikz}

\usepackage{float}

\usepackage{caption}
\usepackage{subcaption}

\usepackage{url}
\usepackage{listings}
\definecolor{codegreen}{rgb}{0,0.6,0}
\definecolor{codegray}{rgb}{0.5,0.5,0.5}
\definecolor{codepurple}{rgb}{0.58,0,0.82}
\definecolor{backcolour}{rgb}{0.95,0.95,0.92}
\definecolor{codeblue}{rgb}{0,0,1}
\definecolor{codered}{rgb}{1,0,0}

\lstdefinestyle{python}{
    backgroundcolor=\color{backcolour},   
    commentstyle=\color{codegreen},
    keywordstyle=\color{magenta},
    numberstyle=\tiny\color{codegray},
    stringstyle=\color{codepurple},
    basicstyle=\footnotesize,
    breakatwhitespace=false,         
    breaklines=true,                 
    captionpos=b,                    
    keepspaces=true,                 
    numbers=left,                    
    numbersep=5pt,                  
    showspaces=false,                
    showstringspaces=false,
    showtabs=false,                  
    tabsize=2
}

\definecolor{backcolour2}{rgb}{0.92,0.95,0.95}
\definecolor{imadcolor}{rgb}{0.8,0.4,0} 
\definecolor{ldgstscolor}{rgb}{0.4,0,0.8} 

\lstdefinestyle{sass}{
    backgroundcolor=\color{backcolour2},   
    commentstyle=\color{codegreen},
    keywordstyle=\color{codeblue},
    numberstyle=\tiny\color{codegray},
    stringstyle=\color{codered},
    basicstyle=\ttfamily\footnotesize,
    breakatwhitespace=false,         
    breaklines=true,                 
    captionpos=b,                    
    keepspaces=true,                 
    numbers=left,                    
    numbersep=5pt,                  
    showspaces=false,                
    showstringspaces=false,
    showtabs=false,                  
    tabsize=2,
    literate={IMAD}{{\textcolor{imadcolor}{IMAD}}}{4}
    {IADD3}{{\textcolor{imadcolor}{IADD3}}}{4}
    {UIADD3}{{\textcolor{imadcolor}{UIADD3}}}{4}
    {HMMA}{{\textcolor{imadcolor}{HMMA}}}{4}
    {LDGSTS}{{\textcolor{ldgstscolor}{LDGSTS}}}{6}
    {STG.E}{{\textcolor{ldgstscolor}{STG.E}}}{6}
    {MOV}{{\textcolor{imadcolor}{MOV}}}{4}
}

\usepackage{multicol}
\definecolor{keywordcolor}{rgb}{0.13,0.29,0.53}
\definecolor{commentcolor}{rgb}{0,0.5,0}

\lstdefinelanguage{PTX}{
    morekeywords={mul, lo, s64, shl, b64, add, s32, selp, b32, cp, async, cg, shared, global, commit, group},
    morecomment=[l]{//},
    keywordstyle=\color{keywordcolor}\bfseries,
    commentstyle=\color{commentcolor}\itshape,
}

\lstdefinestyle{myPTXstyle}{
    language=PTX,
    basicstyle=\footnotesize\ttfamily, 
    breaklines=true, 
    columns=fullflexible, 
    numbers=left, 
    stepnumber=1, 
    frame=single, 
    keepspaces=true, 
    captionpos=b, 
    showstringspaces=false, 
}

\usepackage{enumitem}
\setlist[itemize]{align=parleft,left=0pt..1em}

\usepackage[title]{appendix}

\AtBeginDocument{%
  }

\setcopyright{cc}
\setcctype{by}
\copyrightyear{2025}
\acmYear{2025}
\acmConference[CGO '25]{Proceedings of the 23rd ACM/IEEE International Symposium on Code Generation and Optimization}{March 01--05, 2025}{Las Vegas, NV, USA}
\acmBooktitle{Proceedings of the 23rd ACM/IEEE International Symposium on Code Generation and Optimization (CGO '25), March 01--05, 2025, Las Vegas, NV, USA}
\acmDOI{10.1145/3696443.3708943}
\acmISBN{979-8-4007-1275-3/25/03}

\begin{document}

\title{CuAsmRL: Optimizing GPU SASS Schedules via Deep Reinforcement Learning}

\author{Guoliang He}
\orcid{0009-0008-9713-1678}
\affiliation{%
  \institution{University of Cambridge}
  \city{Cambridge}
  \country{United Kingdom}
}
\email{gh512@cam.ac.uk}

\author{Eiko Yoneki}
\orcid{0000-0002-5552-4536}
\affiliation{%
  \institution{University of Cambridge}
  \city{Cambridge}
  \country{United Kingdom}
}
\email{eiko.yoneki@cl.cam.ac.uk}


\begin{abstract}
  Large language models (LLMs) are remarked by their substantial computational requirements. To mitigate the cost, researchers develop specialized CUDA kernels, which often fuse several tensor operations to maximize the utilization of GPUs as much as possible. However, those specialized kernels may still leave performance on the table as CUDA assembly experts show that manual optimization of GPU SASS schedules can lead to better performance, and trial-and-error is largely employed to manually find the best GPU SASS schedules.

  In this work, we employ an automatic approach to optimize GPU SASS schedules, which thus can be integrated into existing compiler frameworks. The key to automatic optimization is training an RL agent to mimic how human experts perform manual scheduling. To this end, we formulate an assembly game, where RL agents can play to find the best GPU SASS schedules. The assembly game starts from a \textit{-O3} optimized SASS schedule, and the RL agents can iteratively apply actions to mutate the current schedules. Positive rewards are generated if the mutated schedules get higher throughput by executing on GPUs. Experiments show that CuAsmRL can further improve the performance of existing specialized CUDA kernels transparently by up to $26\%$, and on average $9\%$. Moreover, it is used as a tool to reveal potential optimization moves learned automatically.
\end{abstract}

\ccsdesc[500]{Computing methodologies~Massively parallel algorithms}
\ccsdesc[500]{Computing methodologies~Machine learning algorithms}

\keywords{GPU Instruction Scheduling, Reinforcement Learning}


\maketitle

\section{Introduction}

LLMs are transformer-based deep neural networks (DNNs) consisting of many layers of self-attention \cite{attn} and linear projections. Since their appearance, state-of-the-art performance has been achieved across various domains, such as image generation \cite{sora} and natural language processing \cite{llama}. To date, OpenAI \cite{sam_twitter, chatgpt} announces more than 100 billion words are generated every day. As such, LLMs have become a significant workload in the deep learning community and have gathered much attention. 
 
However, training and serving LLMs are computationally expensive because they typically consist of multiple layers of transformer backbone, which is of billions of parameters. As a result, researchers have developed specialized CUDA kernels to accelerate LLM computation, instead of relying on high-level language to generate CUDA kernels. For example, fused attention (flash-attention) \cite{flash_attn} is developed such that the attention computation achieves better utilization of the shared memory of NVIDIA GPUs. Fused feed-forward is a kernel implementation that fuses multiple operators for LLAMA \cite{llama}, and root-mean-square layer normalization is a popular layer normalization operator for transformers \cite{rmsnorm}. We observe that those works are typically implemented by handwritten hardware-efficient codes, i.e. CUDA kernels for NVIDIA GPUs, for the flexibility and efficiency of hardware-vendor-provided programming models. 
 
In this work, we investigate the possibility of further improving the handwritten kernels by exploring optimization at a lower level, i.e. hardware native assembly. Specifically, we focus on NVIDIA CUDA kernels. Optimizing at a lower level allows us to further optimize existing specialized CUDA kernels and this approach has been employed by previous works \cite{opt_conv_gpu,cuda_maxas}, which show that manual optimization of GPU-native assembly schedules can lead to better performance. However, trial-and-error is suggested to manually find the best GPU SASS schedules, which is a tedious process even for CUDA experts, and cannot keep up with the development of new deep learning operators. Moreover, manual optimization cannot be integrated into existing compilation pipelines.
 
We propose CuAsmRL, an automatic optimizer for optimizing NVIDIA GPU SASS schedules. The idea of automatic optimization is achieved by training an RL agent, which mimics how human experts perform manual scheduling, to learn to find the optimized SASS schedule. To the best of our knowledge, we are the first to formulate the optimization of SASS schedules as an assembly game. 

Being able to automatically optimize the SASS schedules enables us to integrate CuAsmRL into OpenAI Triton \cite{triton}, an MLIR-based compiler for writing GPU kernels. Therefore, it first uses an autotuner to find the optimal kernel configurations, and then reuses the compilation pipeline of Triton but intercepts the generated \textit{cubin}, which is then disassembled to SASS instructions, performs optimization and finally assembles back to an optimized \textit{cubin}.



By evaluating on characteristic LLMs kernels, we find CuAsmRL automatically discovered better schedules than the \textit{-O3} SASS schedule, which leads to $1.09$x of geometric mean throughput improvement. As this optimization takes place at a lower level, it is transparent to CUDA kernel developers. Given LLM training and serving can easily consume millions of GPU hours, we expect this kernel-level improvement to be significant. 

In summary, this paper makes the following contributions:

\begin{itemize}
    \item We formulate optimizing SASS schedules as an assembly game, and we implement CuAsmRL, an automatic optimizer for optimizing NVIDIA GPU SASS schedules.
    \item We integrate CuAsmRL into an existing compiler framework, OpenAI Triton, as a SASS-to-SASS optimizer, and it is transparent to CUDA kernel developers.
    \item Our evaluation shows that representative specialized kernels for LLMs can be further accelerated by up to $26\%$ and on average $9\%$ on Ampere GPUs.
    \item We demonstrate CuAsmRL can be used as a tool to reveal optimization moves learned automatically, which can bring new insights into the optimization of SASS instructions.
\end{itemize}

\section{Background and motivations}
 
\subsection{Programming GPUs and compiling CUDA kernels}
 
 
GPUs are hardware accelerators that can perform highly parallel computation and therefore tensor operations can be executed efficiently. To program GPUs, programmers must follow the programming model provided by CUDA \cite{cuda}. Conceptually, a CUDA kernel consists of a grid of thread blocks running concurrently, and inside each thread block are multiple threads. Each thread block is mapped to a GPU steaming multiprocessor and is executed individually and in parallel.

\begin{figure}[htb]
     \centering
     \includegraphics[width=\linewidth]{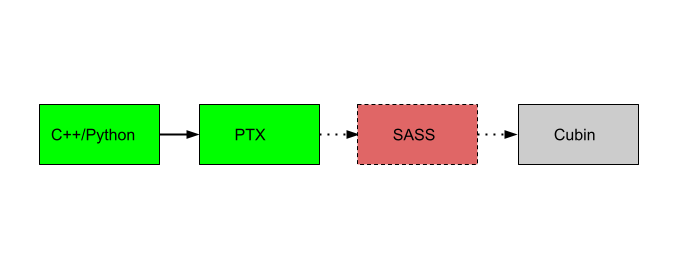}
     \vspace*{-15mm}
     \caption{CUDA compilation as documented by NVIDIA \cite{cuda_compiler}. C++/Python and PTX are highlighted in green, indicating they are the common programming interfaces. SASS is a GPUs-native assembly and is highlighted in red, meaning it is undocumented. Cubin is an executable binary and is in gray.}
     \label{fig: Cuda compilation}

\end{figure}

CUDA kernel developers often program in a high-level programming language, such as C++ or Python, and then compilers compile the kernel code to device code. In the case of C++, the compilation is done by NVIDIA's compiler (\textit{NVCC}), while for Python, Triton \cite{triton} can be used. The compilation process has several stages: first, the code is compiled to PTX, which is an intermediate language that is GPU-architecture independent \cite{cuda_ptx}. Note that one can also directly embed PTX when programming with a high-level programming language.


Then, the PTX codes are compiled to SASS, which is only possible through NVIDIA's proprietary compiler \textit{ptxas} \cite{cuda_ptx}. SASS is a native assembly language to the target GPU. That is, the SASS is specific to the target GPU's architecture. In this work, we limit our discussion to Ampere GPUs. While the corresponding SASS codes of a CUDA kernel are obtainable by utilizing the CUDA binary utilities \cite{cuda_sass}, the instruction set is only vaguely documented. As a result, the lowering and optimization at this stage are unknown and inaccessible.

Finally, the SASS codes are assembled into binary code (\textit{cubin}) that can be executed directly on the GPU. The overall compilation process is shown in Figure \ref{fig: Cuda compilation}.

\subsection{Optimizing GPUs SASS instructions} 
\label{sec. related work}

While there have been extensive works on optimization of CUDA kernels at the C++/Python level, such as memory access \cite{mem_coalecs}, and load balancing \cite{gpu_opt_servey}, there is much less work on optimizing GPU SASS schedules. This is mostly because SASS is closed-source and a lack of official assemblers. However, as SASS is at a lower level in the compilation pipeline, optimization that takes place at this level can be transparent, and all existing specialized CUDA kernels can be beneficial by optimizing their SASS instructions.

Moreover, the open-source community has been able to develop customized assemblers, therefore enabling the optimization at the SASS level. For example, \textit{MaxAs} \cite{cuda_maxas} is the first work on decoding CUDA binary and assembling SASS for early generations of GPUs. After that, assemblers for newer GPU architectures such as \textit{TuringAs} \cite{opt_conv_gpu} and \textit{Cuasm} \cite{cuda_cuasm} have been developed, which enables researchers to optimize their GPU SASS instructions. In the following sections, we first discuss the structure of SASS instructions and then talk about the methodology of optimization employed by those works.

\subsection{Parsing SASS instructions}
\label{sec. sass}

A typical SASS instruction is shown below, which consists of several fields, a control code, an opcode, and operands.

\vspace*{2mm}
\begin{verbatim}
[B------:R-:W2:Y:S02] LDG.E R0, [R2.64];
\end{verbatim}
\vspace*{2mm}

The control code is enclosed by square brackets and is separated into multiple fields by colons \cite{cuda_cuasm}. The first field is the wait barrier mask, if any of the bits are set, the instruction is stalled until the bit is clear. The second and third fields are read and write barrier masks. In this case, this instruction sets the write-barrier to 2, which means a later instruction using the data in $R0$ is stalled until $R0$ is ready. The fourth field is the yield flag, which is believed to be used for load balancing \cite{opt_conv_gpu}. Finally, the last field is the stall count, which indicates how many cycles to stall the current instruction before issuing the next one. 

The opcode is only vaguely documented on the official website \cite{cuda_sass}, and in this case \textbf{LDG} stands for loading data from global memory. The operands consist of registers and memory addresses. For a more systematic decoding of the SASS instructions, it is recommended to read a prior work \cite{opt_conv_gpu}.


\subsubsection{Fixed and variable latency instructions}

SASS instructions can be classified as fixed latency instruction and variable latency instruction. Fixed latency instructions, such as \textbf{IADD3} and \textbf{FFMA}, are usually mathematical operations and take a fixed number of cycles to execute, while variable latency instructions, such as \textbf{LDG.E} (loading data from global memory), take a variable number of cycles to execute due to the deep hierarchy of GPU memory system, which consists of L1, L2 caches and global memory. As such, it is impossible to know the cycles needed for accessing data in advance. Moreover, since the \textit{Kepler} GPU architecture \cite{cuda_kepler}, the execution of instructions is static, indicating the compiler must prevent data hazards. Therefore, the control code associated with the instruction stalls the instruction until the data is ready.

For example, the above \textbf{LDG.E} instruction has variable latency, and therefore its control code indicates setting the $2$nd write barrier. The user of $R0$ will be stalled until the barrier is clear.


\subsection{Latency hiding} 
\label{sec. latency_hiding}

The stall of execution due to resolving data dependencies introduces latency, and GPUs have two mechanisms to perform latency hiding. As soon as a warp performs a long-latency operation, the latency is hidden by the hardware by either \textit{1)} switching to the next eligible warps or \textit{2)} scheduling the next independent instruction. The two forms are referred to as thread-level parallelism (TLP) and instruction-level parallelism (ILP) respectively \cite{gpu_opt_servey}.

Prior works \cite{opt_conv_gpu,opt_gemm_gpu,cuda_maxas,microarch} show a methodology for hiding memory access latency by manually reordering SASS instructions, which overlaps the memory load/store and computation instructions as much as possible. This is a form of improving ILP because the instruction execution pipeline is less likely to stall. While the GPU could switch to another eligible warp, i.e. TLP, the number of eligible warps may run out because it depends on the algorithms as well as kernel configurations such as the tile sizes as well as the usage of registers, and the stall may eventually slow down the overall execution.


As a result, there have been attempts to hide latency by manually interleaving memory load/store and compute instructions. In \textit{MaxAs} \cite{cuda_maxas}, a trial-and-error strategy is employed. In \textit{TuringAs} \cite{opt_conv_gpu}, a profiling-guided strategy is employed.

\subsection{Reinforcement learning}

Reinforcement learning (RL) is a group of algorithms designed to solve sequential decision-making problems by iteratively acting in the environment and learning from the consequences. To apply RL, users typically need to define the optimization problem as a Markov decision process (MDP), which consists of the state space, the action space, and the reward function \cite{rlBook}. RL is an intelligent learning algorithm under the sequential decision-making framework for its optimization towards a long-term reward. 

\begin{equation}
   \pi = \argmax_{\pi}  \mathbf{E}[\sum_{t=0}^{\infty} \gamma^{t}r_t | s_0]
   \label{eq. rl objective}
\end{equation}

In recent years, deep RL  refers to RL algorithms that use deep neural networks to learn the optimal policy from a given MDP. The advantages of applying RL are that it can learn complex and dynamic decision-making problems with little human intervention. Moreover, the optimization objectives of RL are typically long-term rewards, meaning RL agents can learn to tolerate short-term losses and maximize long-term gains. Therefore, deep RL has been successfully applied to a wide range of domains, including video games \cite{atari, Go}, robotic control tasks \cite{openai2019solving}, data center power management, and device placement \cite{addanki2019placeto, mirhoseini2018hierarchical}. 

%

\subsection{Motivations}



We observe that existing optimization on SASS schedules requires enormous manual work and is error-prone. Firstly, each kernel consists of several thousand lines of SASS instructions, and optimizable patterns must be identified manually. Secondly, the dependencies between SASS instructions must be preserved carefully. Moreover, if any of the input data types or the kernel configurations change, the SASS instructions are completely different and must be re-optimized. Finally, manual scheduling is not integrable to existing compiler frameworks unless it can be automated. 



We aim to apply RL to bridge this gap. This is because instruction interleaving can be formulated as a discrete optimization problem, where RL can learn to take a sequence of actions to maximize the long-term reward. Furthermore, RL-based optimization is automated, meaning we can incorporate the optimizer into an existing compiler framework. In this way, CUDA kernels that are compiled by the compiler can be optimized by RL agents automatically with minimal human intervention.

\section{CuAsmRL}

\subsection{Hierarchical search space}
\label{sec. overview}

\begin{figure}[htb!]
     \centering
      \includegraphics[width=0.6\linewidth]{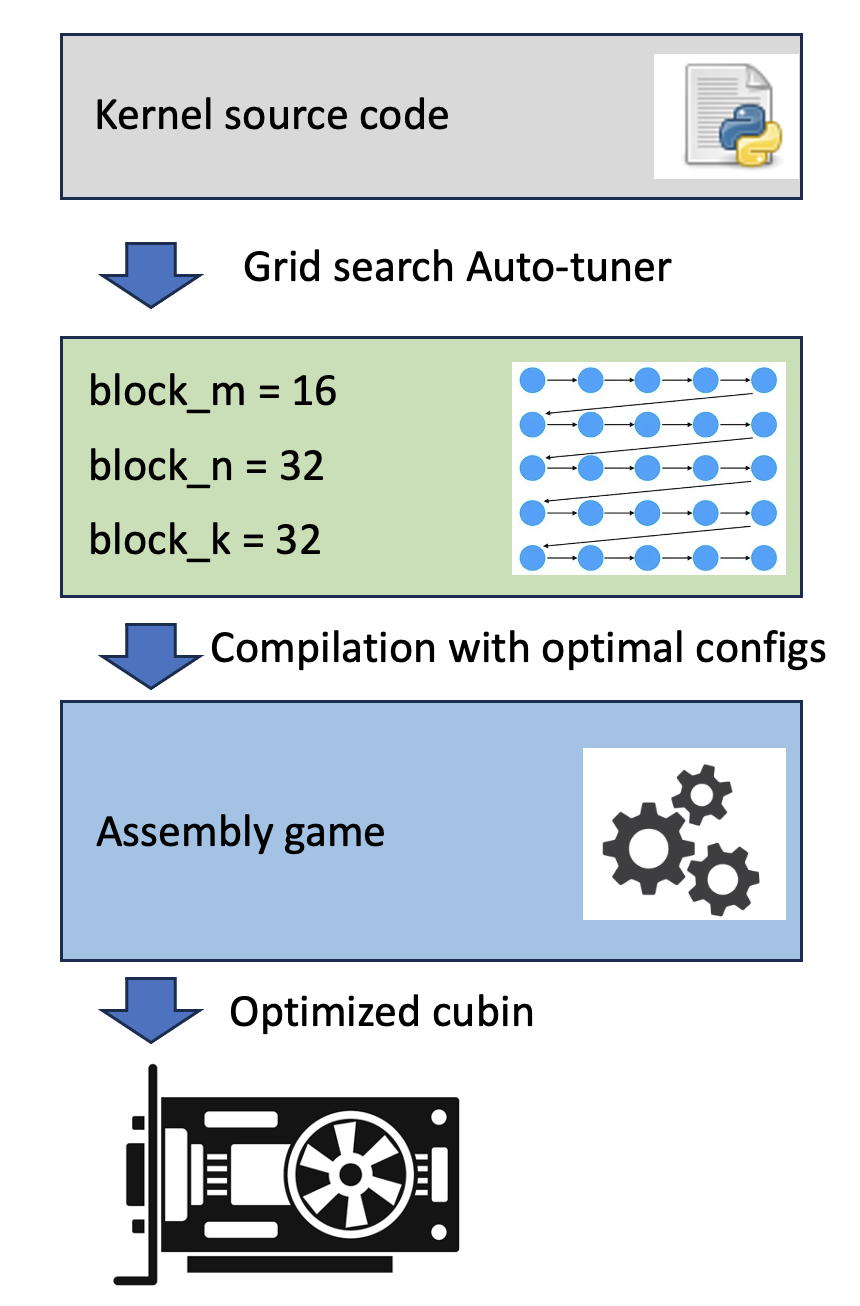}
      \caption{Overall workflow of CuAsmRL. CuAsmRL takes as input the source code targeting Triton's programming interface. Then it uses an autotuner to enumerate and find the optimal kernel configurations. Then the code is compiled with the optimal kernel configurations via Triton's compilation pipeline. Finally, an RL agent is trained to play the assembly game to optimize the SASS schedules, which outputs an optimized \textit{cubin}.}
      \label{fig: CuAsmRL}
\end{figure}

In this section, we give an overview of CuAsmRL, an automatic optimizer for the SASS schedules. In the following sections, we first introduce the hierarchical search space, and then we formulate the SASS scheduling as a reinforcement learning problem.

Figure \ref{fig: CuAsmRL} shows the overall architecture of CuAsmRL. Because of its integration with Triton, it takes as input the kernel source code written to target Triton's programming interface and uses the existing autotuning pipeline to find the optimal kernel configurations. Then the compilation pipeline compiles the kernel to generate \textit{cubin}, which is disassembled to SASS instructions with an official tool \cite{cuda_sass}. Finally, an RL agent is trained to play the assembly game to optimize the SASS schedules, which outputs an optimized \textit{cubin}. 

The autotuner is essential, as the kernel configurations such as the tile sizes can lead to up to $2$x throughput difference and completely different SASS instructions, which results in a different SASS schedule for the RL agent to optimize. As such, we perform a hierarchical search, which first finds the optimal kernel configurations and then optimizes the SASS schedule based on the optimal kernel configurations. The autotuner employs a grid search-like strategy, which enumerates user-provided kernel configurations, compiles with the kernel configurations, measures the execution throughput on the target GPU, and greedily selects as well as caches the optimal set of kernel configurations. The measurement is performed by taking the average of $100$ repeated execution, preceded by $100$ warm-up iterations.

\subsection{Pre-game static analysis}
\label{sec. static analysis}

CuAsmRL has a parser to decode SASS instructions. Besides simply separating an instruction into different parts, such as control codes, opcodes etc., and storing to a data structure, it also expands the operands. General-purpose registers are 32-bit, and we find that if they are suffixed with \textit{.64}, it indicates the adjacent registers are involved in the operation. This can be verified by constructing a microbenchmark, deliberately contaminating the adjacent register, and then comparing the output to the expected value.

As this pattern is commonly observed in memory instructions, we expand the operand with the adjacent registers to retrieve the correct dependencies. We use the following algorithm to determine the adjacent register:
  
\begin{equation} 
  \begin{split}
  \textit{base} & = (\textit{No. of reg}) / 2 \\
  \textit{mod} & =   (\textit{No. of reg}) \% 2 \\ 
  \textit{flip} & =   1 - \textit{mod} \\
  \textit{adj.reg} & = \textit{base}*2 + \textit{flip}
  \end{split}
\end{equation}

Before initializing the assembly game, several analysis passes are run through the assembly file to perform static analysis.

\begin{itemize}

  \item An analysis pass tries to record every memory instruction if it consumes the output of a fixed latency instruction in the same basic block. For every memory instruction, the analysis pass looks up the assignment of its operand registers by scanning its preceding instructions. If a label is encountered first, the analysis pass aborts and adds the current memory instruction to a denylist. Otherwise, the accumulated stall count between the use-definition instruction pair is recorded. If the stall count of a fixed latency instruction is already recorded, either from micro-benchmarks (\S \ref{sec. stall count table}) or from a previously inferred value, we take the minimum one. 

  The analysis takes place within the same basic block because we do not allow reordering instructions across labels (\S \ref{sec: action space}). We find this analysis pass is very powerful in practice. For example, running this pass on one kernel can infer the stall count of \textbf{IADD3.X} is $5$, which is only $1$ cycle away from the result of the microbenchmark. The slight overestimation is safe, and because of the original schedule is always valid, the inferred value would be either overestimated or exact. In the future, instead of performing the manual micro-benchmarking, we can potentially run this pass on a large amount of SASS kernel codes and build a stall count look-up table automatically. For example, with every release of the CUDA toolkit, lots of kernels in shared libraries (\textit{libcu*.so}) can be dumped and analyzed.
  

  \item An analysis pass prepares for embedding (\S \ref{sec: state space}). For example, it builds a table that maps operand registers to integers. Also, because SASS instructions have a variable number of operands, we record the maximum number of operands in the assembly file. Instructions with fewer operands are padded with dummy values ($-1$) during embedding.

  \item An analysis pass counts the number of memory instructions in the SASS file except for those in the denylist, which is used to define the action space, detailed in \S \ref{sec: action space}. 
\end{itemize}


\subsection{Reinforcement learning}

\begin{figure}[htb!]
  \centering
   \includegraphics[width=0.8\linewidth]{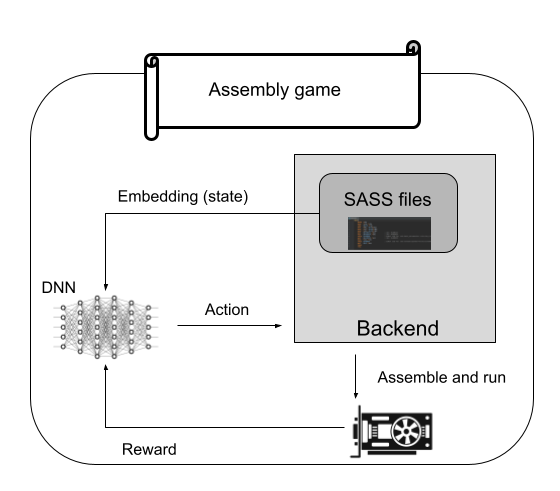}
   \caption{Assembly Game. At each iteration $i$, the SASS file is embedded, and the embedding is fed to the RL agent as state $S_i$. The RL agent is represented by a deep neural network. The output of the RL agent is an action $A_i$ that changes the SASS file. Then the mutated SASS file is assembled and sent to execution on the target GPU. A reward $R_i$ is sent back to the agent and the mutated SASS file is transitioned to the next state $S_{i+1}$.}
   \label{fig: assembly game}
\end{figure}

Having analyzed the disassembled SASS instructions, an RL agent is trained to play the assembly game to optimize the SASS schedules. The assembly game is iterative - at each iteration, the RL agent perceives the current SASS schedule (the state) and then takes an action, which changes the SASS schedule. The mutated SASS file is assembled and sent to execution on a GPU, which returns a reward to the agent. This is illustrated by Figure \ref{fig: assembly game}. To formulate the assembly game, we define action space, state space and reward function respectively in the following sections.

\subsection{State space} 
\label{sec: state space}

To represent the SASS schedule as such it is consumable by the RL agent, we embed the SASS instructions.

\begin{figure}[htb!]
  \centering
   \includegraphics[width=\linewidth]{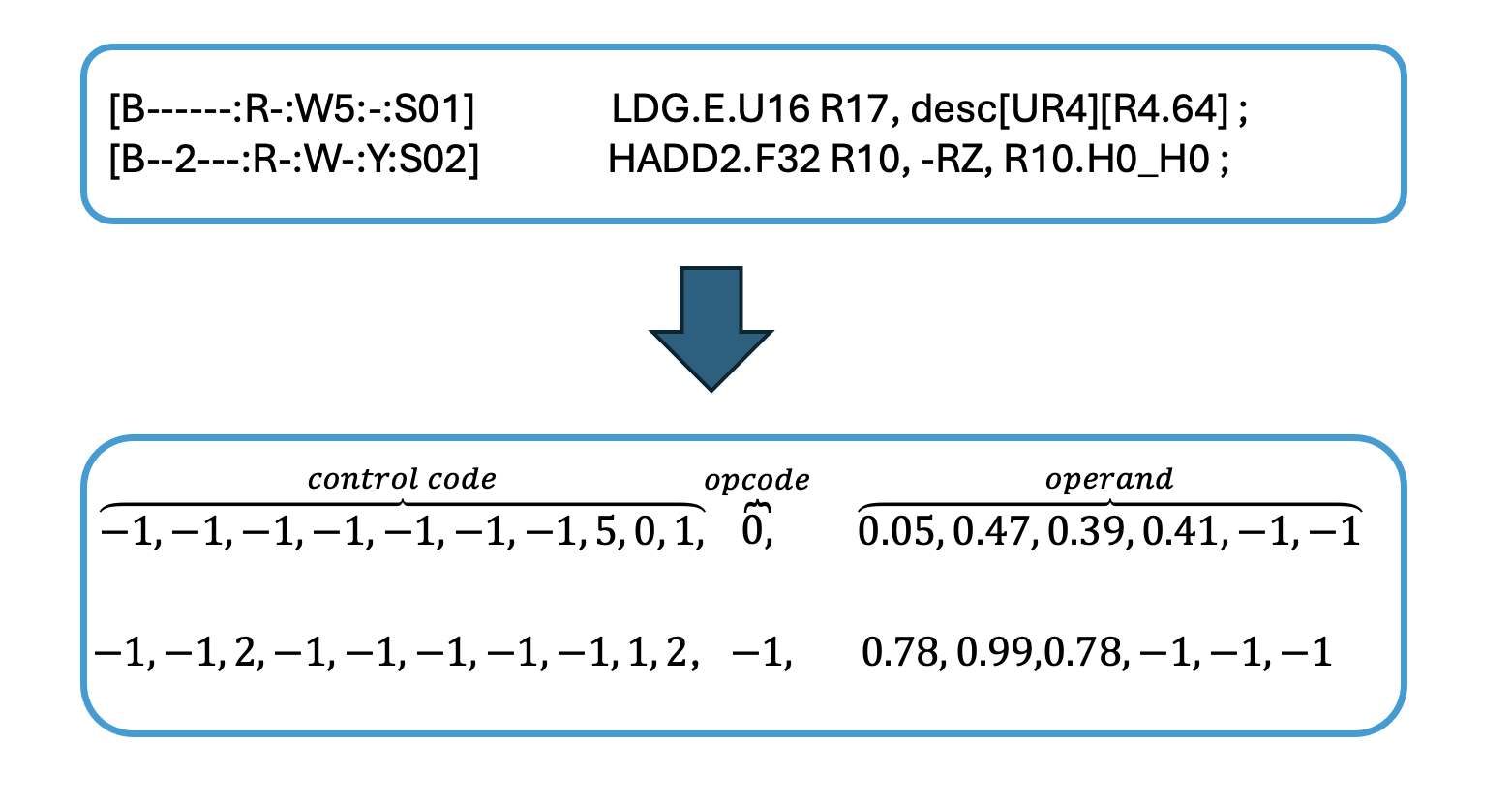}
   \caption{Embedding. Different fields of SASS instruction such as control code, opcode, and operands are embedded individually and then concatenated to a vector. Dummy values (-1) are used for the absent fields and operand padding. Different vectors are concatenated in a row-wise fashion. The final embedding of the assembly file becomes a matrix, which represents the state of the SASS file.}
   \label{fig: embedding}
\end{figure}

Recall that a typical SASS instruction consists of a control code, an opcode, and operands as shown in \S \ref{sec. sass}, we embed each field individually and concatenate the embeddings. For example, the read and write barrier can take any integer from $0$ to $5$, and so do their embeddings. If the barrier is absent, a $-1$ is filled. For opcode, we only classify whether it is a memory instruction or non-memory instruction. The pre-game analysis passes have extracted the memory instructions from the SASS file. For non-memory instruction, a $-1$ is used. For operands, we convert the memory locations to their indices in the memory table, which is built by the pre-game analysis pass, and then we normalize those indices by dividing them by the total number of memory locations. $-1$ will be padded until the number of operands matches the maximum of the number of operands in the SASS file because SASS instruction has a variable number of operands. An example of embedding SASS instructions is shown in Figure \ref{fig: embedding}.

Therefore, after the embedding, the state representation of one SASS instruction is a vector, and the assembly file becomes a matrix by concatenating the instruction vectors in a row-wise fashion.

\subsection{Action space} 
\label{sec: action space}

With the definition of the state space, we then define the action space. Considering the process in which experts interleave the compute instruction and memory instructions, we want to allow our agent to have the same flexibility. As such, we allow the agent to select an instruction and swap it with the instruction above or below. We think this resembles how experts perform the interleaving, which is illustrated by Figure \ref{fig. action example}.

\begin{figure}[ht]
  \centering

  \begin{minipage}{0.4\textwidth}
    \begin{lstlisting}[caption=Before, style=sass]
IMAD.WIDE R14, R84, R8, c[0x0][0x160] ;
LDGSTS.E.BYPASS.LTC128B.128 [R74], desc[UR18][R18.64], P4 ;
      \end{lstlisting}
  \end{minipage}

  \begin{minipage}{0.1\textwidth}
      \centering
      \tikz{\draw[thick,->] (0,0) -- (0, -1);} 
  \end{minipage}\hfill

  \begin{minipage}{0.4\textwidth}
    \begin{lstlisting}[caption=After, style=sass]
LDGSTS.E.BYPASS.LTC128B.128 [R74], desc[UR18][R18.64], P4 ;
IMAD.WIDE R14, R84, R8, c[0x0][0x160] ;
      \end{lstlisting}
  \end{minipage}

  \caption{An example of an action, which reorders the SASS instructions.}
  \label{fig. action example}
\end{figure}

However, allowing each instruction to be reordered introduces a massive action space, as a kernel can have thousands of lines of SASS instructions. Considering the latency hiding process is mostly about placing the memory load/store instruction at a better location, we can only explore a small subset of the action, which prunes the action space. Specifically, we only allow the RL agent to pick memory load/store instructions, such as \textbf{LDG}, \textbf{LDGSTS}, and \textbf{STG}, whose indices are recorded by the pre-game analysis pass and are dynamically updated at each iteration. The RL agent outputs a discrete number, which is mapped to the index of an instruction and the direction of the reordering. The RL agent has a Convolutional Neural Network (CNN) for encoding the state representation, followed by an MLP layer to output the probability of each action. 

It is also crucial to preserve data dependencies during the reordering process, as violations can result in incorrect results. To this end, we employ action masking to filter out potential invalid actions. We have the following dependencies to consider:

\begin{itemize}
  \item Register dependencies: the users of a register cannot be reordered such that it is before the assignment.

  \item Barrier dependencies: the read and write barrier cannot be reordered before any of their setters. For example, if an instruction waits for the $2$nd barrier, then it cannot be reordered such that it comes before the setter of the $2$nd barrier. This is achieved by comparing the control codes of adjacent instructions.

  \item Stall count dependencies: the fixed latency instruction resolves the dependencies by stalling the instruction for a fixed number of cycles, which is indicated by the stall count number. As the unmodified SASS instructions are scheduled by NVIDIA's proprietary compiler, the dependencies are always satisfied. While this number is not publically released, we obtain the stall count values either through microbenchmarks (\S \ref{sec. stall count table}) or through the analysis pass (\S \ref{sec. static analysis}). If a memory instruction uses registers assigned by a fixed latency instruction with unknown stall count, the analysis pass adds the memory instruction to a denylist, whose instructions are always masked out. Otherwise, we check its preceding and following instructions to see whether a reordering may cause a potential violation.

  For example, the action masking algorithm for checking whether stall count is satisfied if moving a memory instruction up is shown by Algorithm \ref{algo. action masking}. It accumulates the stall count and compares it with the minimum stall count. If the accumulated stall count is less than the minimum, the action is masked. 



  \item Additional dependencies: there are additional dependencies to be considered. For example, we find that when a sequence of \textbf{LDGSTS} writes to consecutive memory addresses offset by a register, reordering any of them will cause an error. It is likely associated with hardware design that transfers data from global memory to shared memory for Ampere GPUs, and we have to identify them manually, because of the lack of publically available data. We also prevent instructions from moving across labels or any barrier/synchronization instructions, so instructions are only rescheduled within the same basic block. A list of barrier/synchronization instructions is shown in the official specification\footnote{\url{https://docs.nvidia.com/cuda/cuda-binary-utilities/index.html}}.
  
  Additional dependencies are represented as heuristic rules and are hard-coded. Any action that may lead to violation of the heuristic rules are masked out. As the LLM domain is characterized by a few kernels evaluated in \S \ref{sec. evaluation}, we find the current heuristic rules set sufficient in the domain. In \S \ref{sec. auto_discovery}, we also manually verify the reordering process step-by-step for the optimized kernels.
\end{itemize}

\begin{algorithm}
  \caption{Algorithm for masking stall count}
  \begin{algorithmic}[1]
  \State Initialize mask = 1, accum = 0 
  \State Initialize cur = index of current SASS instruction

    \vspace*{1mm}

    \While {true}
    \State inst\_to\_check = cur - i
    \State stall\_count = get\_stall\_count(inst\_to\_check)
    \State accum += stall\_count

    \vspace*{1mm}
    \If {is\_user(inst\_to\_check, cur)}
    \State min\_st = get\_min\_stall\_count(inst\_to\_check)
      \If {accum $<$ min\_st}
        \State mask = 0
      \EndIf
      \State Break
    \EndIf
    \EndWhile

  \State \textbf{return} mask
  \end{algorithmic}
  \label{algo. action masking}
  \end{algorithm}

With those dependencies to consider, we generate a mask for each action, which is dynamically updated at each iteration and for each action. If an action may lead to any potential violation of the dependencies, the masking number is $0$ which assigns an impossible probability to the action. If no actions are available, the episode is terminated immediately.

\subsection{Reward function}

Obtaining the feedback signal is the most important component as it directly guides the RL to explore good schedules. In this work, we mostly care about the runtime of the optimized CUDA kernels, and therefore we must measure the runtime after an action is applied. 

Specifically, we use CUDA events to measure the kernel execution time. We follow a standard approach by first warming up the GPU for $100$ iterations and then repeating $100$ iterations to measure the elapsed time \cite{cuda_event}. L2 caches are cleared between iterations to get an accurate measurement. The average execution time is returned as the feedback signal. We observe the standard deviation of two individual measurements is typically within $1\%$ of each other. We use the following formula to obtain the reward:


\begin{equation}
  R_i = \frac{T_{i-1} - T_i}{T_0} * 100
\end{equation}


Where $T_0$ is the initial runtime, $T_i$ is the runtime after an action is applied, $T_{i-1}$ is the runtime before the action is applied, and $100$ is the scaling factor. Intuitively, this gives positive feedback if the action decreases the runtime, and negative feedback otherwise. According to the optimizing objective function as shown in Equation \ref{eq. rl objective}, the RL agent learns a policy, represented by its policy neural network, that aims to maximize the cumulative reward which leads to reduce the total kernel execution time. This objective function also encourages the RL agent to tolerate short-term losses if actions can bring long-term rewards.

\subsection{RL algorithm}


By default, CuAsmRL has a reference implementation of the proximal policy optimization algorithm (PPO) \cite{ppo}, and we use the same set of hyperparameters for all cases, as fine-tuning RL's hyperparameters towards a specific case is very computationally expensive. The default hyperparameters are taken from a study \cite{the37implementation}, which performs large-scale case study across various domains, and summarizes an empirically good set of hyperparameters. In \S \ref{sec. rl stats}, we also investigate the sensitivity of the algorithm under different hyperparameter settings. 

We modify the implementation to use a CNN to encode the embedding of the assembly file and then use an actor-critic policy gradient algorithm to learn the optimal policy. As the reordering process is encapsulated in the environment transition, which followed the standardized \textit{Gym} interface \cite{openaigym}, we expect changes to future RL algorithms to be easy. Training statistics such as episodic rewards and the loss of the RL agents are logged and the agent's weight is checkpointed periodically.

\section{Implementation}
\label{sec. impl}

\subsection{Integration to Triton}



We choose to integrate CuAsmRL with OpenAI Triton \cite{triton}, which is a compiler for writing GPU kernels. Triton allows users to write kernel codes in Python syntax and then just-in-time compile to either NVIDIA GPUs or AMD GPUs. Moreover, Triton is also the default backend of Pytorch \cite{pt2}, one of the most popular deep-learning frameworks. By integrating with Triton, we hope our work can be beneficial to the deep-learning community directly. 


The syntax of writing kernels in Triton is shown by the Listing \ref{lst. triton}.


\begin{lstlisting}[language=Python, caption=Example Triton kernel codes, style=python,label=lst. triton]
  @triton.jit
  def matmul(x_ptr, y_ptr, out_ptr):
    ...
\end{lstlisting}

CuAsmRL reuses Triton's compilation pipeline but extends the autotuner and intercepts the compiled \textit{cubin}. It then disassembles the \textit{cubin} into SASS and extracts the kernel section consisting of SASS schedules while keeping the other meta-information intact. This is important as the meta-information such as the symbol tables and the ELF format must be preserved. Then it trains RL agents to optimize the kernel section and substitutes the kernel section with the optimized \textit{cubin}. To apply CuAsmRL's optimization, users simply need to change one line in the Triton code as shown in Listing \ref{lst. cuasmrl example}.

\begin{lstlisting}[language=Python, caption=CuAsmRL example, style=python,label=lst. cuasmrl example]
  @cuasmrl.jit(ret_ptr=1)
  def matmul(x_ptr, y_ptr, out_ptr):
    ...
\end{lstlisting}

Where the \textit{ret\_ptr} is the index to the output buffer and can be used for probabilistic testing. Probabilistic testing generates randomized inputs and reference outputs and then compared with the output of the program. We use probabilistic testing as a sanity check, and we also manually verify each step of the optimized kernels, detailed in \S \ref{sec. auto_discovery}. Formal verification methods cannot apply to SASS sequences due to the lack of official semantics, and bitwise enumeration of the test inputs is computationally intractable, as kernels typically process large amount of input data. Optionally, users may add more arguments to specify the hyperparameters of the RL agents, such as the learning rate, the batch size for training etc. 

\subsection{Workflow}

As training RL agents is a time-consuming process, we expect users to employ an offline search and deploy-time lookup workflow. This is also because more training budget allocated to the RL agent may lead to better exploration of the action space, which leads to better performance. Listing \ref{lst: Cuasmrl usages} shows how to invoke the optimization of CuAsmRL and the deployment with an optimized \textit{cubin}.

\begin{center}
\begin{minipage}{0.45\textwidth}
\begin{lstlisting}[language=Python, caption=CuAsmRL invoke optimization and deployment example, style=python, label=lst: Cuasmrl usages]
# invoke optimization
matmul(x_ptr, y_ptr, out_ptr)
# deploy
matmul(x_ptr, y_ptr, out_ptr, load_dir='path-to-cubin')
\end{lstlisting}
\end{minipage}
\end{center}

After writing a kernel, users should invoke CuAsmRL which performs hierarchical optimization. Then the best optimized \textit{cubin} found throughout the assembly game is written to the file system, prefixed by GPU type, workload type etc., as the key to lookup. At deployment, the key should be passed in, and it invokes a lookup process instead of training, which finds the best \textit{cubin} and loads it into Triton. Therefore, there will be no runtime overhead but just offline search time. We observe the training time of RL agents is typically less than $5$ hours, which is a one-time cost and is negligible because LLM training and serving can consume millions of GPU hours.





\subsection{Stall count table}
\label{sec. stall count table}

CuAsmRL has a built-in table that maps the names of common fixed-latency instructions to their corresponding stall counts. This table is obtained by performing microbenchmarking, and it is be used by the action masking detailed in section \S\ref{sec: action space}. The table is presented in Table \ref{tab. stall counts}. It covers the common integer operations, because they are frequently involved in address calculation, and thus their outputs are often consumed by later memory instructions. 

\begin{table}[h]
  \centering
  \footnotesize
  \caption{Fixed-latency instructions and their stall counts on A100 GPU.}
  \begin{tabular}{|l|c|c|}
      \hline
       Instructions & Stall counts (cycles)  \\
      \hline
       IADD3, IMAD.IADD, IADD3.X, MOV, IABS & 4 \\
        IMAD,FADD, HADD2, IMNMX, SEL, LEA &  \\
      \hline
       IMAD.WIDE, IMAD.WIDE.U32 & 5 \\
      \hline
  \end{tabular}
  \label{tab. stall counts}
\end{table}

We describe how the micro-benchmarking is performed. Unlike a prior work \cite{demystifyingampere} that performs micro-benchmarking in PTX for Ampere GPUs, we directly program SASS instructions. This allows us to construct use-definition instruction pairs to accurately determine the stall counts for fixed-latency instructions. The methodology is employed by previous works on dissecting Volta and Turing GPUs \cite{dissect_volta,dissect_turing}. We start by writing a simple CUDA kernel, compile and dump its SASS instructions, and based on which we program SASS instructions. For example, Listing \ref{lst. sass microbenchmark} shows the microbenchmark for the \textbf{MOV} instruction.


\begin{center}
\begin{minipage}{0.45\textwidth}
\begin{lstlisting}[caption=dependency-based SASS microbenchmark, style=sass, label=lst. sass microbenchmark]
[B------:R-:W-:-:S04]MOV R15, 0x1;
[B------:R-:W-:-:S04]STG.E desc[UR4][R4.64], R15;
\end{lstlisting}
\end{minipage}
\end{center}


As the user instruction (line 2) consumes the output of the \textbf{MOV} instruction (line 1) and stores it in global memory, we gradually lower the stall count of the \textbf{MOV} instruction until the output does not match the expected value. The minimum stall count is then the number of cycles needed for the \textbf{MOV} instruction to stall. 

With \textbf{MOV} known, we can control the values held by registers and subsequently construct similar microbenchmarks for other instructions. For instructions that need more stall counts, we insert \textbf{NOP} in between until the output matches the expected value. Those stall count values are then hard-coded in CuAsmRL.

We find that dependency-based micro benchmarking is more accurate than clock-based micro benchmarking, as used by a previous work \cite{demystifyingampere}, which can underestimate the stall count. Considering the clock-based micro benchmarking in Listing \ref{lst. time sass microbenchmark} (control codes are omitted):

\begin{center}
\begin{minipage}{0.45\textwidth}
\begin{lstlisting}[caption=clock-based SASS microbenchmark, style=sass, label=lst. time sass microbenchmark]
CS2R R2, SR_CLOCKLO ;  // t1
// IADD3 sequence...
CS2R R6, SR_CLOCKLO ; // t2
IADD3 R6, P0, -R2, R6, RZ ; // t2 - t1
\end{lstlisting}
\end{minipage}
\end{center}

The measured averaged stall count for the \textbf{IADD3} instruction is $2.6$ cycles if we evaluate the clock, which does not match Table \ref{tab. stall counts}. We think this is because, at the time of the second clock ($t2$), there is no guarantee that all \textbf{IADD3} instructions have finished execution, thus leading to underestimated clock cycles. To mitigate the issue, one would need to construct artificial read/write dependencies of the \textbf{IADD3} sequences and the last timing instruction. This indicates the necessity of utilizing the dependency between SASS instructions to accurately measure the stall count.

\section{Evaluation}
\label{sec. evaluation}
In this section, we aim to evaluate CuAsmRL to answer the following questions:

\begin{figure*}[ht!]
  \centering
  \includegraphics[width=\linewidth]{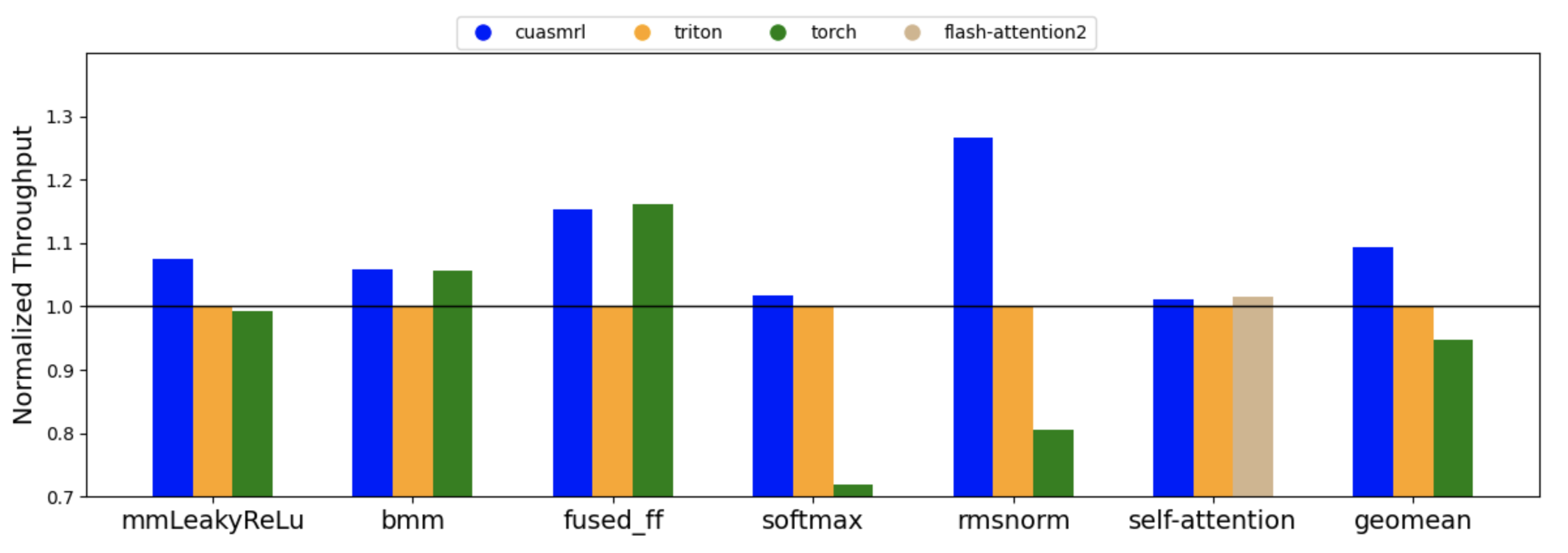}
  \caption{Overall kernel throughput comparison. The throughput of Triton is normalized to $1$, and the others are normalized accordingly. A high value indicates a better performance compared to Triton. \textit{bmm}: batch matrix multiplication, \textit{fused\_ff}: fused feed-forward, \textit{rmsnorm}: root-mean-square layer normalization, \textit{mmLeakyReLu}: matrix multiplication with LeakyReLU.}
  \label{fig: kernel throughput}
\end{figure*}

\begin{itemize}
  \item How much speedup can CuAsmRL achieve transparently over Triton and other baselines?
  \item Is CuAsmRL sensitive to its hyperparameters configurations?
  \item Why is it necessary to optimize at the SASS level, and what are the optimization moves taken by the RL agents to better schedule SASS instructions?
\end{itemize}

\subsection{Experiment setup}


We evaluate CuAsmRL with an NVIDIA A100 80GB PCIe GPU (Ampere architecture). We use the NVIDIA compiler \textit{ptxas} $12.2$ and Triton v$2.1.0$. As CuAsmRL is meant to be a SASS-to-SASS optimizer that further optimizes the best existing SASS schedules and it is integrated into Triton, we compare it to common LLM kernels developed in Triton. Additionally, we construct a Pytorch (v$2.1.2$) baseline by composing Pytorch operations. Pytorch's eager operations dispatch kernels to CuBLAS \cite{cuda_cublas} (v$12.1$) - NVIDIA's high-performance library, which however provides limited customization of fusion. We also construct a Cutlass (v$3.5$) baseline for fused GEMM with LeakyReLU and a flash-attention (v$2.3.3$) baseline for self-attention computation.


To benchmark kernel throughput, we take the average of $5$ runs, each of which uses CUDA events to measure the kernel execution time, by warming up $100$ iterations and repeating $100$ iterations. To study fine-grained kernel metric (\S \ref{sec. breakdown}), we dump the optimized \textit{cubin} to the file system after training, and use Nsight Compute \cite{ncomp}, a kernel profiler, to study the hardware metrics of the optimized kernels from CuAsmRL and Triton respectively. Nsight Compute can be used to extract fine-grained statistics of the optimized kernels with access to NVIDIA's GPU performance counter \cite{nv_perfcounter}.

We choose to evaluate CuAsmRL on representative kernels for LLMs. For example, compute-intensive kernels include fused GEMM and epilogue (Leaky-ReLU), fused feed-forward, batch matrix multiplication and flash-attention \cite{llama,mixtral,attn}, whereas memory-bound kernels include \textit{Rmsnorm} and \textit{Softmax}. Those fused kernels are taken from the Triton repository \cite{triton_repo} and the \textit{Kernl} repository \cite{kernl}. Common kernel sizes and configurations (\textit{float16} data type) are applied. A summary of the evaluated kernels is listed in Table \ref{tab. kernels}. 


\begin{table}[h]
  \centering
  \footnotesize
  \caption{Evaluated Kernels}
  \begin{tabular}{|l|c|c|}
      \hline
      \textbf{Compute-bound}  & \textbf{inputs} & \textbf{configuration}  \\
      \hline
      fused\_ff & B, M, N, K & 1, 512, 512, 2048 \\
      \hline
      mmLeakyReLu & B, M, N, K & 1, 512, 512, 2048\\
      \hline
      bmm & B, M, N, K & 4, 512, 512, 2048 \\
      \hline
      flash-attention & B, n\_head, seq\_len, d\_head & 1, 4, 4096, 32 \\
      \hline
      \hline
      \textbf{Memory-bound}  & \textbf{inputs} & \textbf{configuration}  \\
      \hline
      softmax & n\_rows, n\_cols & 512, 4096 \\
      \hline
      rmsnorm & B, n\_head, seq\_len, d\_head & 1, 32, 4096, 64 \\
      \hline
  \end{tabular}
  \label{tab. kernels}
\end{table}

\subsection{Instruction latency}

We have described our micro-benchmarking approach to measure the stall count of fixed-latency instructions in section \S\ref{sec. stall count table}, and presented the main results in Table \ref{tab. stall counts}. We find common integer operations have a stall count of $4$ cycles, which is similar to the previous Volta and Turing GPUs \cite{dissect_turing}. This may indicate the integer operations unit of GPUs has not changed over the last few generations.

\begin{figure}[htb!]
  \vspace*{-5mm}
  \centering
  \includegraphics[width=0.8\linewidth]{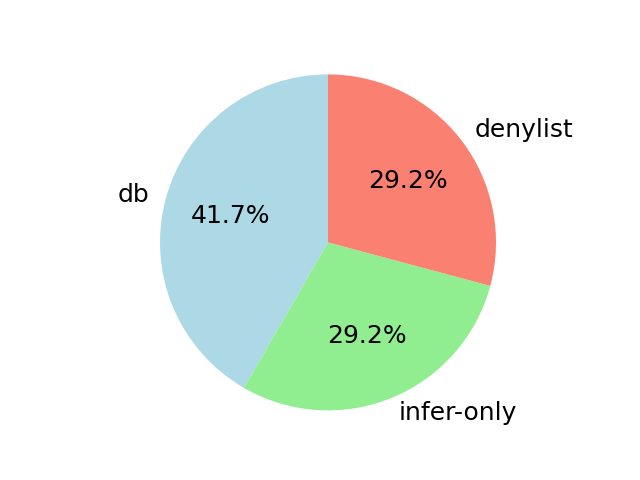}
  \vspace*{-3mm}
  \caption{Percentages of stall count of fixed-latency instructions that are resolved by the built-in stall count table (db), inferred by the analysis pass (infer-only), and deny-listed (not resolved) on average for kernels listed in Table \ref{tab. kernels}.}
  \label{fig. analysis}
\end{figure}

Figure \ref{fig. analysis} shows the percentages of stall count dependencies that are resolved by the looking up the stall count table, inferred, or deny-listed by the analysis pass mentioned in \S \ref{sec. static analysis}. We find on average, $41.7\%$ of stall count dependencies can be resolved by the built-in stall count table. This indicates the effectiveness of Table \ref{tab. stall counts}, as common integer operations are micro-benchmarked, and they are frequently involved in address calculation. On the other hand, as opcode can change behavior by suffixing a modifier, such as \textbf{IMAD}, \textbf{IMAD.MOV} and \textbf{IMAD.WIDE} etc., the analysis pass can further infer $29.2\%$ of the stall count dependencies. If more instruction latency is added to the stall count table, we expect the ratio of \textit{db} can be further improved, however the ratio of denylist will remain the same, as their dependencies must be resolved by crossing basic blocks, which requires control flow analysis for SASS instructions. 

\subsection{Kernel throughput}


Figure \ref{fig: kernel throughput} shows the normalized kernel throughput achieved by CuAsmRL, Triton, and other baselines. CuAsmRL consistently outperforms Triton on all kernels, indicating it is capable of further improving the performance by optimizing the SASS schedules. 

For batch matrix multiplication, fused feed-forward and flash-attention, the kernels from Triton are slower than those from reference implementation (CuBLAS and Flash-Attention2). This is because the reference implementation consists of highly engineered and optimized codes, which requires access to a lower-level programming interface than the one provided by Triton. Nevertheless, CuAsmRL is able to further improve the performance on top of Triton-generated code, matching the reference implementation.


For fused GEMM with LeakyReLU, softmax and root-mean-square layer normalization, Triton is more advantageous than Pytorch, because it can fuse multiple smaller operators into one kernel, instead of composing operations. This indicates the flexibility of Triton's programming interface while achieves comparable performance to reference implementation. Moreover, CuAsmRL can further improve on those kernels, transparently producing $2\%$ to $26\%$ speedup. We also benchmark the Cutlass implementation on fused GEMM with LeakyReLU with the default configuration and find it achieves very limited performance ($10$x less throughput than Triton). We suspect this is due to the suboptimality of the default configuration, and without an autotuner users must invest effort to tune the configurations, such as block sizes, pipelining stages etc.




\subsection{Speedups breakdown analysis}
\label{sec. breakdown}

In this section, we use Nsight Compute to study the fine-grained statistics of the optimized kernels from CuAsmRL and Triton. The compute workload analysis and memory workload analysis reported by Nsight Compute show a detailed analysis of the compute resources utilized by the streaming multiprocessor (SM) as well as memory resources respectively.

\begin{table}[htb!]
  \centering
  \footnotesize
  \caption{Compute and memory workload analysis of fused GEMM with the epilogue.}
  \begin{tabular}{|c|c|c|c|}
    \hline
     && CuAsmRL & Triton \\
    \hline
    Compute & Executed Ipc Active (inst/cycle) & 0.75&0.74 \\
    Resources      & Executed Ipc Elapsed (inst/cycle) & 0.59 & 0.52\\
            & SM Busy  (\%)  & 25.54 & 25.11 \\
    \hline
    Memory & Memory Throughput (GB/s) & 175.71&157.73 \\
     Resources       &  Mem Busy  (\%)   & 45.58 &  40.54\\
            & Max Bandwidth  (\%)  & 42.33 & 37.63 \\
    \hline
  \end{tabular}
  \label{tab. breakdown}
\end{table}
\vspace*{3mm}

As shown by Table \ref{tab. breakdown}, the optimized kernel of fused GEMM with LeakyReLU from CuAsmRL and Triton have negligible differences in utilizing computer resources because the instruction per clock (IPC) achieves similar values. Also, the SM busy time is similar in both CuAsmRL and Triton, indicating the amount of computation is similar. On the other hand, the memory throughput of CuAsmRL is $175$GB/s, $11\%$ higher than that of Triton. This can be attributed to a higher memory busy percentage, $45.58\%$ over $40.54\%$. This indicates the optimized schedule better utilities the memory resources while keeping the same utilization of the compute resources.

\begin{figure*}[htb!]
  \centering
  \subfloat{{\includegraphics[width=0.47\linewidth]{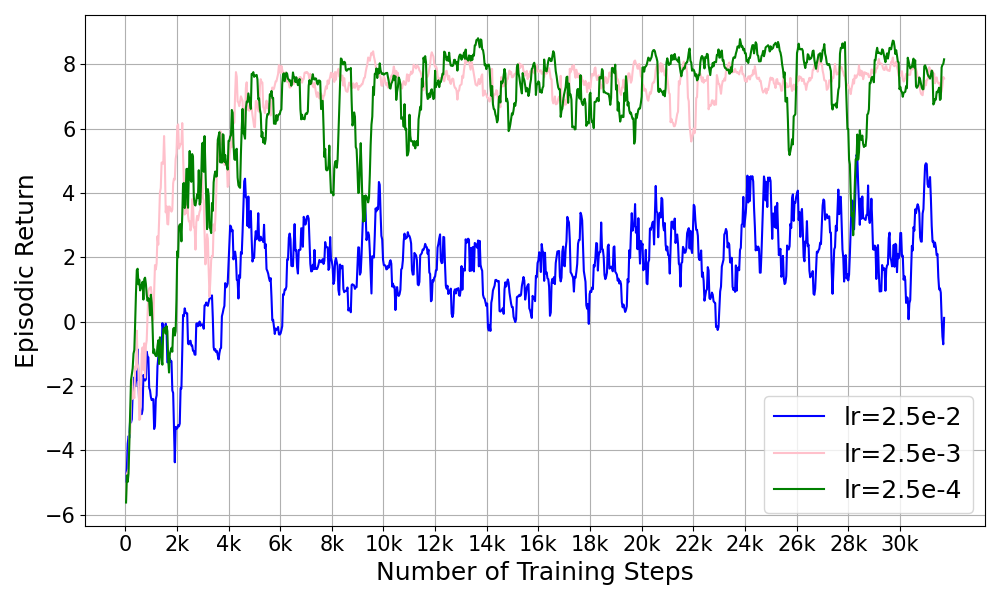} }
  \label{fig. lr}
  }%
  \qquad
  \subfloat{{\includegraphics[width=0.47\linewidth]{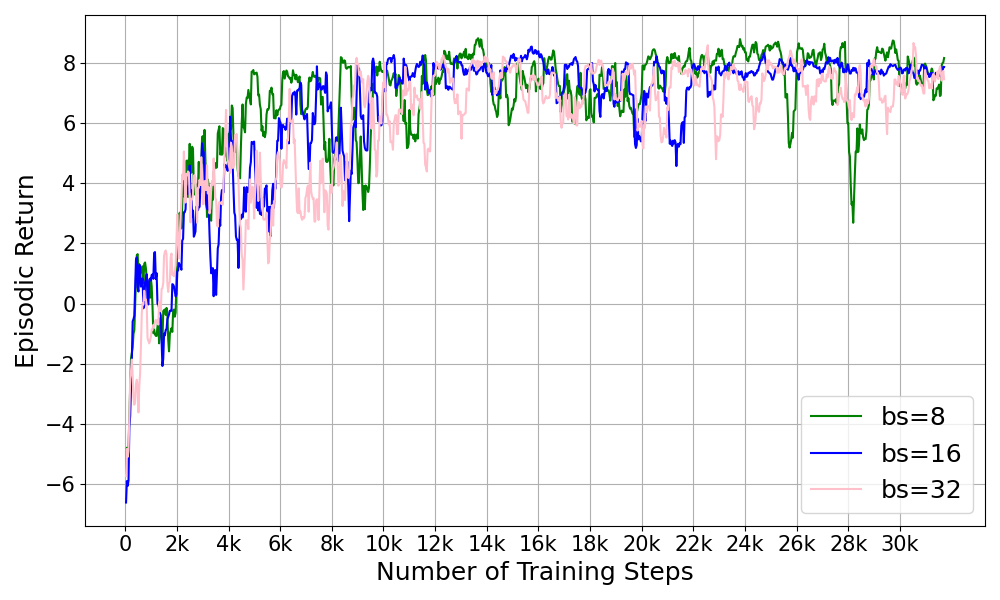} }
  \label{fig. bs}
  }%
  \caption{Episodic returns for different hyperparameter settings. The green line is the default setting.}
  \label{fig: hyperparams}%
\end{figure*}

We also provide the memory chart from Nsight Compute in Figure \ref{fig. mt-cuasmrl} and Figure \ref{fig. mt-tt} in the Appendix \ref{appendix. memory chart}. It can be observed that the memory throughput from global memory to shared memory is significantly improved by CuAsmRL. In \S \ref{sec. auto_discovery}, we show more details of how the memory throughput is improved by showing the optimization moves performed by the RL agent.

\subsection{RL training statistics}
\label{sec. rl stats}

Figure \ref{fig: hyperparams} studies the sensitivity of the RL agent to different hyperparameters when optimizing fused GEMM with LeakyReLU. Two of the most significant hyperparameters, i.e. learning rate and training batch size are swept. We can observe that under the default hyperparameters setting, the RL agent consistently converges to achieve the best episodic return, indicating the robustness of the setting. Note that the default hyperparameters setting come from a work which performs large-scale case study across various domains \cite{the37implementation}.


Figure \ref{fig: training} in the Appendix \ref{appendix. rl stats} shows an example of time series plots during the training process. Specifically, the approximated KL divergence measures the distance between the updated policy network and the old network, whereas policy entropy measures the uncertainty of the policy network. Both metrics decrease over training steps, indicating the policy network of the RL agent gradually converges, and thus each update round is less and less diverted.

\subsection{Necessity for SASS-level optimization}
\label{sec. necessity}

In this section, we investigate the necessity of performing optimization at the SASS level. Specifically, we compare the PTX code and SASS instructions taken from the same CUDA kernel. Note that the SASS presented in this section is specific to the NVIDIA Ampere GPUs. The comparison is shown by the Listing in Appendix \ref{appendix. comparison}.

Considering the PTX code snippet in Listing \ref{lst:ptx_example}, where a sequence of operations is performed to calculate the address and to load data from global memory to shared memory. The corresponding SASS is listed in Listing \ref{lst. triton_sass_example}. Note that the consecutive \textbf{cp.async} (in PTX) is translated to \textbf{LDGSTS} (native to Ampere GPUs) and interleaved with address calculation automatically (\textbf{IMDA} instructions) by the compiler (\textit{ptxas}'s $-O3$ optimization). This illustrates the necessity of SASS-level optimization because higher-level codes such as PTX are compiled and transformed into hardware-native assembly (SASS), and reordering at the PTX level is not able to control the specific memory load/store SASS instructions. In \S \ref{sec. auto_discovery}, we show the exact placement of memory load/store in the SASS schedule is crucial to obtaining a better performance. 

%

\subsection{Automatic Discovery of optimization moves}
\label{sec. auto_discovery}

We can trace the actions taken by the RL agents to discover the optimized SASS schedules and observe which reordering sequence is the most significant. CuAsmRL has a flag that can be toggled by users to trigger the inference mode and a pre-train agent weight file must be provided. The inference process can be seeded, so it is deterministic and can be reproduced. To the best of our knowledge, the optimization moves presented in this section are published for the first time on Ampere GPUs and are learned by the RL agents automatically. Control codes are ignored for simplicity, and some opcodes are simplified. The optimization moves are illustrated by Figure \ref{fig. fused GEMM trace} and \ref{fig. BMM trace}. 

\subsubsection{Fused GEMM with LeakyReLU}

\begin{figure}[htb!]
  \vspace*{-3mm}
  \centering
  \includegraphics[width=0.8\linewidth]{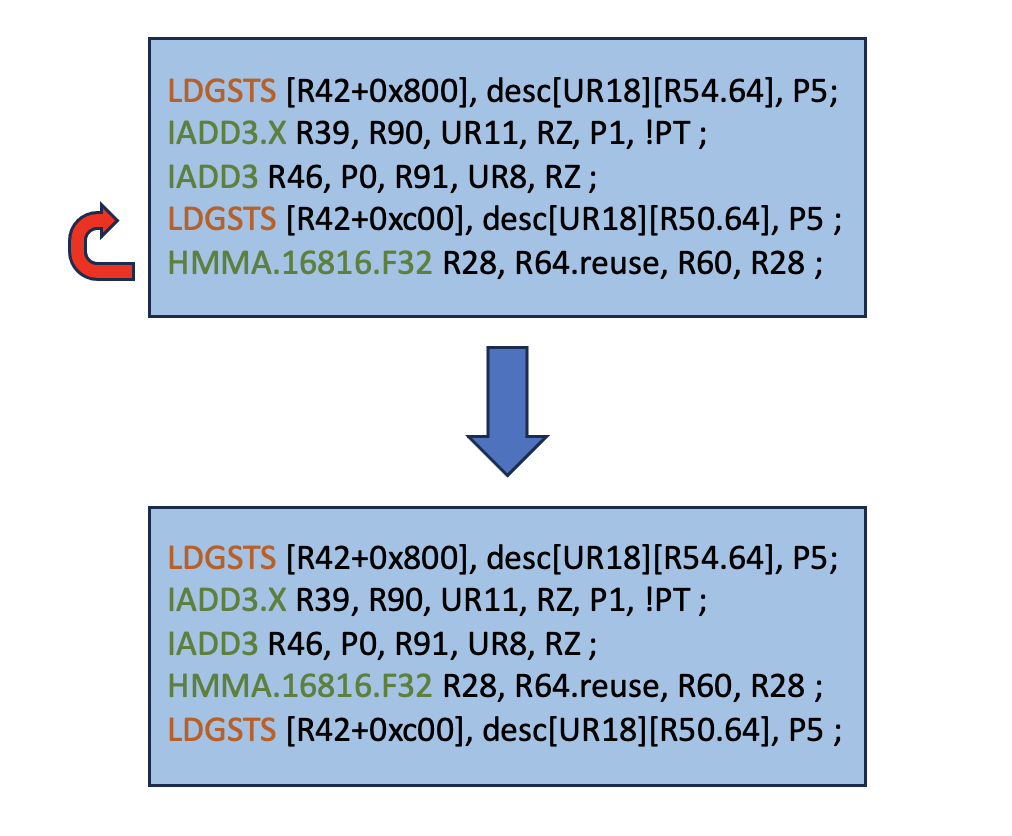}
  \caption{A reordering for fused GEMM and the epilogue. Scheduling the \textbf{HMMA} instruction before the \textbf{LDGSTS} instruction achieves better performance.}
  \label{fig. fused GEMM trace}
\end{figure}

Figure \ref{fig. fused GEMM trace} shows the most significant reordering for the fused GEMM with LeakyReLU. By just reordering the \textbf{HMMA} and \textbf{LDGSTS} instructions, we observe $7\%$ improvement of the kernel throughput. 



By further inspecting the SASS sequence, we suspect the optimization is to do with the \textit{.reuse} flag of the operand register. Indeed, if we manually remove the flag from the original SASS schedule, we observe no performance degradation, whereas if removing it from the optimized schedule, the performance gain is lost. As pointed out by \textit{Maxas}, the \textit{.reuse} flag hints to reuse the operand cache, which helps mitigate the register bank conflict\cite{cuda_maxas}. We hypothesize what happens is that, the compiler attempts to reuse the operand cache when scheduling instructions, however at runtime, the warp scheduler performs a switch at the second \textbf{LDGSTS} due to long latency or insufficient load/store units (TLP), which invalidates the operand cache. This would explain why removing the flag from the original SASS schedule causes no performance degradation. The optimized SASS schedule, on the other hand, is able to reuse the operand cache, and if we remove the flag, the performance gain is lost. The phenomenon indicates the interplay between ILP and TLP, and by perform rescheduling, we can better hide latency.

\subsubsection{Batch matrix multiplication}

Another optimization move that is observed both for fused GEMM LeakyReLU and batch matrix multiplication is shown in Figure \ref{fig. BMM trace} in Appendix \ref{appendix. opt-move}. The \textbf{LDS} instructions are predicated by the guard register \textit{@!PT}, which is always evaluated as false. According to the official guide, instructions with the guard predicate control the conditional execution of the instruction \cite{cuda_ptx}. In this case, CuAsmRL learns to schedule the \textbf{LDGSTS} instruction earlier than the \textbf{LDS} instruction, which is not executed due to its guard predicate.

We also observe that the RL agent becomes lingering after it applies all the necessary optimization moves, by repeatedly moving an instruction up and then down, until the end of the episode. The length of the episode is $32$ and is a hyperparameter for RL training. We find this number is sufficient for our cases, and if the lingering behavior is not observed for other kernels, users may consider increasing the length of the episode and re-start training.

\section{Related works}

%

\subsection{Manual scheduling of SASS instructions}


Prior works on optimizing SASS instructions such as \textit{KeplerAs} \cite{microarch}, \textit{MaxAs} \cite{cuda_maxas} and \textit{TuringAs} \cite{opt_conv_gpu} involves comprehensive profiling of the GPU memory systems and instruction latency, which is then leveraged by CUDA experts to better place the memory load/store. While the approach works well, it is not scalable as each developed CUDA kernel requires a manual optimization process and GPUs are becoming heterogeneous, i.e. different GPUs present unique characteristics even if they belong to the same generation. As CuAsmRL is the first data-driven approach to automate the SASS rescheduling process, it can be applied to a wide spectrum of CUDA kernels. Other instruction scheduling algorithms exist as compiler passes \cite{is_gpu, opt_occ_ilp}, which however cannot be applied to NVIDIA GPUs.

\subsection{Reinforcement learning for compiler optimization}

In recent years, due to the potential of solving NP-hardness problems, RL has been widely applied to optimizing compilers. For example, there have been attempts to tune compiler flags \cite{autophase}, IR transformation \cite{xrlflow}, and even super-optimization \cite{alphadev}. A particularly related work applies RL to schedule instruction in basic blocks \cite{rl_bb_sched}. While those works have covered various aspects in compiler optimization, none of them applies RL to scheduling instructions for GPUs, which have unique challenges for having very different memory hierarchies and computation units compared to CPUs. As such, CuAsmRL differs from prior works in considering the characteristics of GPUs when scheduling instructions, and it is equipped with state-of-the-art RL algorithms.

\section{Limitation and future work}

Applying CuAsmRL to optimize kernels from other domains may require more additional dependencies other than the ones mentioned in \S \ref{sec: action space}, due to the lack of publically available data. Thus, users are required to manually verify the optimized kernels as in \S \ref{sec. auto_discovery}.

Another limitation of CuAsmRL is that it relies on executing GPU kernels on GPUs to obtain the feedback signal and computes the reward function. This means $200$ kernel execution is required every step and typically $15$k steps are needed to train a good policy as shown by Figure \ref{fig: hyperparams}. Thus, a cost model that can approximate the kernel execution time will significantly reduce the training cost. However, the cost model will be challenging because the data of SASS are not publically available.

Given our reordering formulation, it is also possible to apply other search algorithms, such as evolutionary search, to reschedule instructions. Evolutionary search does not need training, however it may converge to local minima and thus has performance degradation. We choose RL for its state-of-the-art performance across various domains, and its potential to generalize to unseen SASS schedules. However, to achieve generalization, we need to pre-train the RL agent across SASS schedules from different CUDA kernels in the future. In that case, the pre-trained RL agent can be incorporated as a regular compiler pass, without needing to spend hours on training from scratch for every CUDA kernel.



\section{Conclusion}


We introduce CuAsmRL, an automatic optimizer for GPU SASS schedules. CuAsmRL performs optimization at the GPU native assembly level, and it can be integrated into existing compiler frameworks while being transparent to CUDA kernel developers. We show that the common kernels in LLMs can be improved by up to $26\%$ and on average $9\%$, and we show the robustness of its hyperparameters and enabling to discover new optimization moves.



\begin{acks}
We gratefully thank the anonymous reviewers and our shepherd for their suggestions and feedback that helped improve this paper. Thanks to Da Yan for sharing his insights and suggestions on decoding the SASS instructions.
\end{acks}

\begin{appendices}

\section{Artifact Appendix}

\subsection{Abstract}
This artifact appendix helps the readers run the artifact and reproduce main results of CuAsmRL. Figure \ref{fig: kernel throughput} consists of $6$ common LLMs kernels to be evaluated, and we provide scripts that run training to optimize each of them, followed by an inference process to load the optimized kernels and then compared against other baselines. The artifact has been uploaded\cite{artifact2025}.

\subsection{Artifact Checklist}

\begin{itemize}
  \item \textbf{Compilation}: NVIDIA CUDA compiler (nvcc)
  \item \textbf{Run-time environment}: Linux Ubuntu 22.04+
  \item \textbf{Hardware}: NVIDIA A100-80GB-PCIe and Intel(R) Xeon(R) Silver 4210R CPU @ 2.40GHz
  \item \textbf{Metric}: kernel throughput
  \item \textbf{How much disk space required (approximately)?} 50 GB
  \item \textbf{How much time is needed to prepare workflow (approximately)?} 1 hour.
  \item \textbf{How much time is needed to complete experiments (approximately)?:} 50 hours. 
  \item \textbf{Code licenses (if publicly available)?:} Apache License v2.0.
  \item \textbf{Publicly available?:} Yes.
  \item \textbf{Archived (provide DOI)?} \url{https://doi.org/10.5281/zenodo.14058861} \cite{artifact2025}
\end{itemize}

\subsection{Description}

\subsubsection{How to access.}
The source code can be downloaded from either the Zenodo archive (\url{https://doi.org/10.5281/zenodo.14058861}) or GitHub repository (\url{https://github.com/hgl71964/cuasmrl/tree/reproduce})

\subsubsection{Hardware and software dependencies.}
The artifact is evaluated in a virtual machine environment running Linux Ubuntu 22.04, with an NVIDIA A100-80GB-PCIe GPU, as well as the following software dependencies:

\begin{itemize}
  \item NVIDIA ptxas 12.2
  \item Triton v2.1.0
  \item Pytorch v2.1.2
  \item NVIDIA CuBLAS library v12.1
  \item Cutlass v3.5
  \item flash-attention v2.3.3
  \item CuAssembler
\end{itemize}

\subsection{Installation} 

See install from source section in README.

\subsection{Experiment workflow} 

For each kernel, CuAsmRL first invokes a RL training process and then the optimized kernels are cached to deploy and use directly. To invoke training and optimization, execute the \textit{benchmarks/train.sh} script. After training is completed, execute the \textit{benchmarks/inference.sh} script to run benchmark against other baselines (Triton, Torch). 

\subsection{Evaluation and expected results}

To get stabilized results, each benchmark is run $5$ times, which is done by the inference script, and the average value should be similar to Figure \ref{fig: kernel throughput}. Each run should output the following format, which represents the measured kernel throughput. 

\begin{table}[htb!]
  \centering
  \footnotesize
  \begin{tabular}{c|c|c}
    \hline
     Torch & CuAsmRL & Triton \\
    \hline
    a & b & c \\
    \hline
  \end{tabular}
\end{table}

\end{appendices}

\bibliographystyle{ACM-Reference-Format}
\bibliography{sample-base}


\clearpage
\onecolumn
\begin{appendices}


\newpage
\section{Memory chart}
\label{appendix. memory chart}

\begin{figure*}[htb!]
  \centering
  \includegraphics[width=\linewidth]{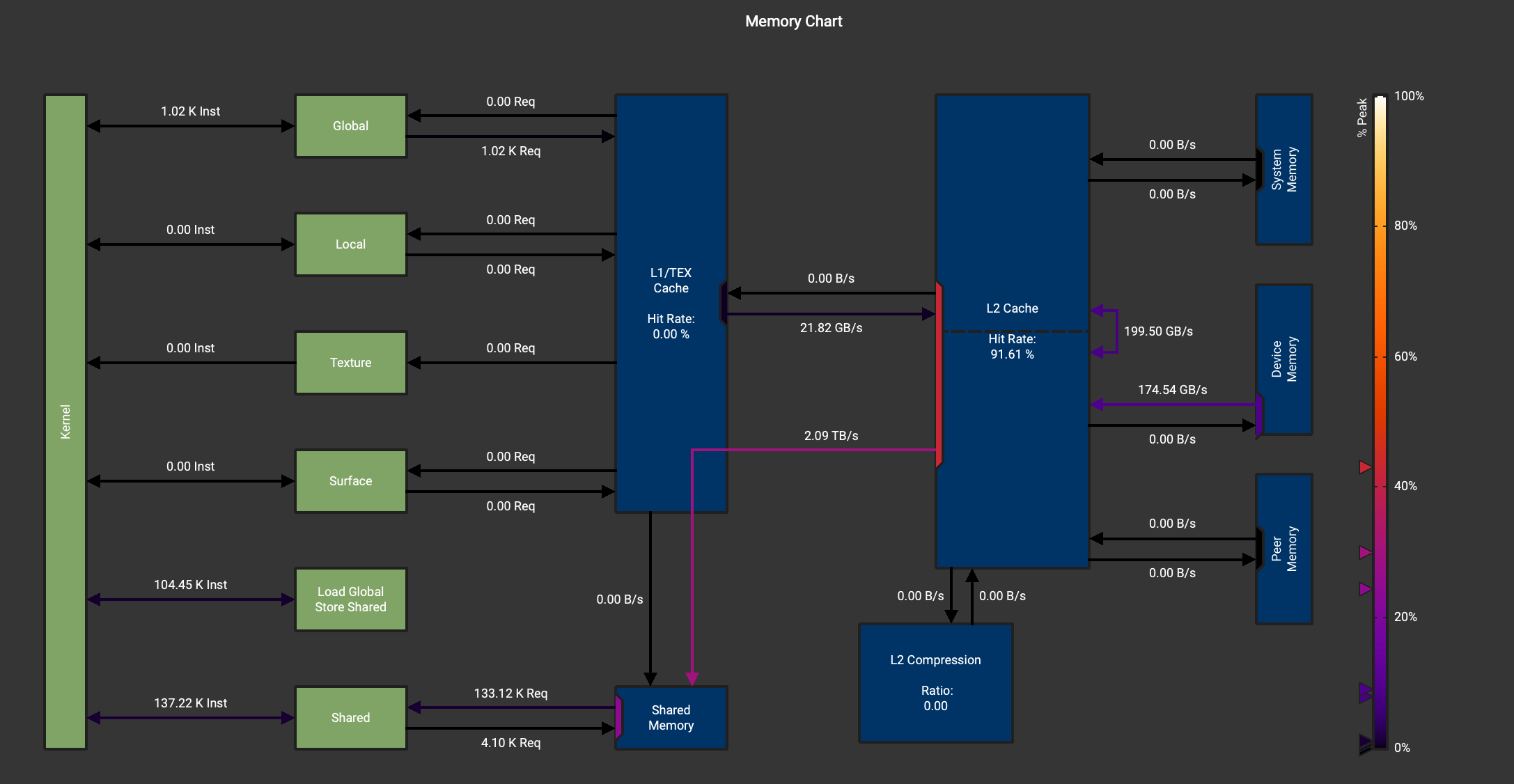}
  \caption{Memory chart for fused GEMM LeakyReLU optimized by CuAsmRL.}
  \label{fig. mt-cuasmrl}
\end{figure*}

\begin{figure*}[htb!]
  \centering
  \includegraphics[width=\linewidth]{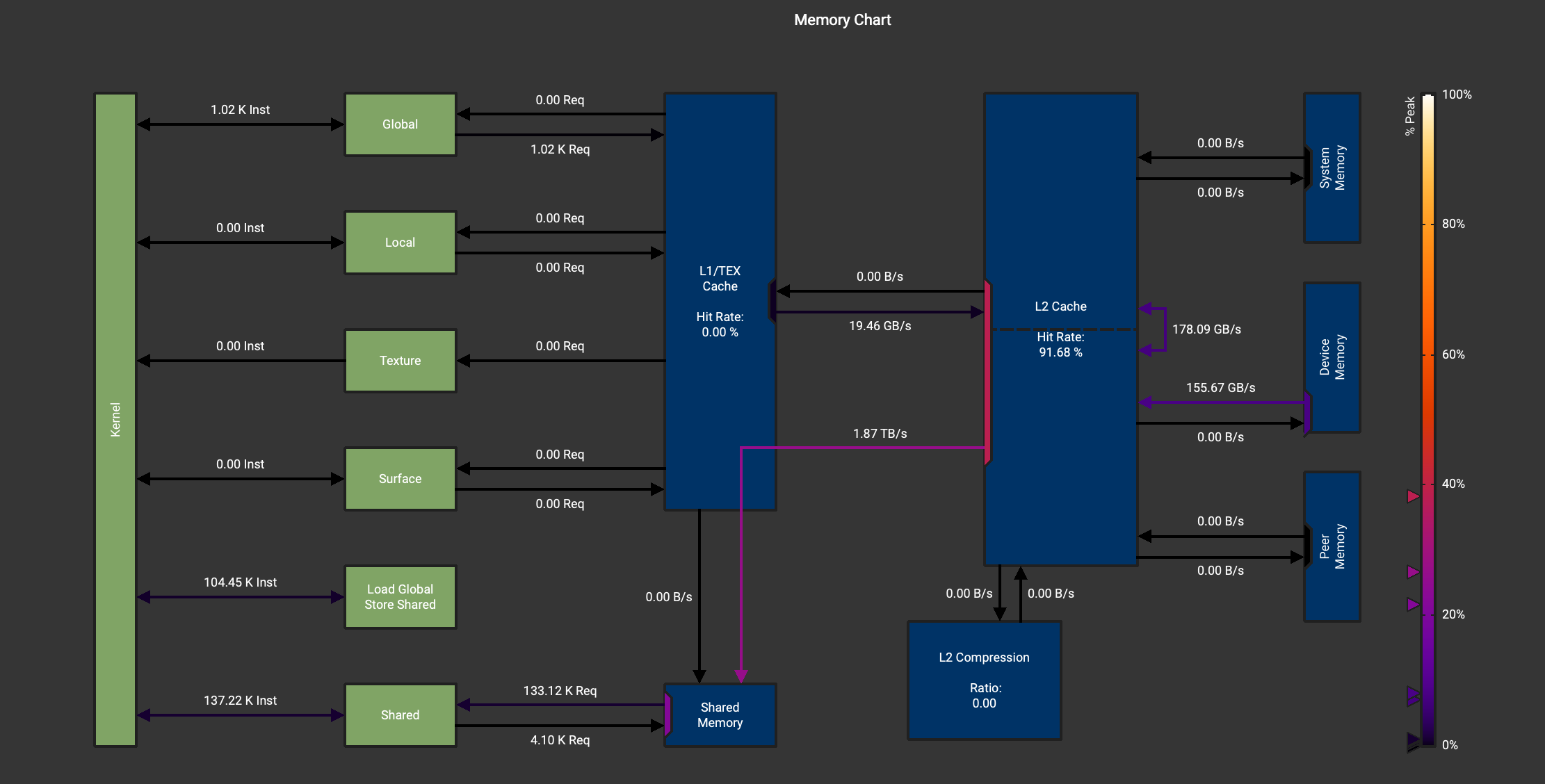}
  \caption{Memory chart for fused GEMM LeakyReLU optimized by Triton.}
  \label{fig. mt-tt}
\end{figure*}

\newpage
\section{RL training statistics}
\label{appendix. rl stats}


\begin{figure*}[htb!]
  \centering
  \subfloat[\centering approximate KL divergence]{{\includegraphics[width=5cm]{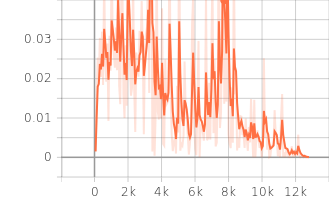} }
  \label{fig. finetune return}
  }%
  \qquad
  \subfloat[\centering policy entropy]{{\includegraphics[width=5cm]{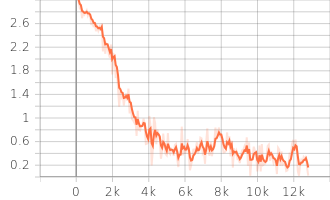} }
  \label{fig. finetune entropy}
  }%
  \caption{Time series plots of training statistics during the assembly game.}
  \label{fig: training}%
\end{figure*}




\section{Comparison of PTX and SASS}
\label{appendix. comparison}

%
%
%

\centering
\begin{minipage}{0.8\textwidth}
\begin{lstlisting}[style=myPTXstyle, caption={PTX}, label=lst:ptx_example]
  add.s32      %r121, %r204, 18432;
  add.s32      %r123, %r204, 20480;
  add.s32      %r125, %r204, 22528;
  selp.b32     %r120, 16, 0, %p10;
  cp.async.cg.shared.global [ %r119 + 0 ], [ %rd86 + 0 ], 0x10, %r120;
  cp.async.cg.shared.global [ %r121 + 0 ], [ %rd87 + 0 ], 0x10, %r120;
  cp.async.cg.shared.global [ %r123 + 0 ], [ %rd88 + 0 ], 0x10, %r120;
  cp.async.cg.shared.global [ %r125 + 0 ], [ %rd89 + 0 ], 0x10, %r120;
  cp.async.commit_group ;
\end{lstlisting}
\end{minipage}

\begin{minipage}{0.8\textwidth}
\begin{lstlisting}[caption=SASS, style=sass, label=lst. triton_sass_example]
LDGSTS.E.BYPASS.128 [R219+0x4000], desc[UR16][R10.64], P0 ;
IMAD.WIDE R18, R9, 0x80, R10 ;
LDGSTS.E.BYPASS.128 [R219+0x4800], desc[UR16][R44.64], P0 ;
IMAD.WIDE.U32 R16, R222, 0x2, R64 ;
LDGSTS.E.BYPASS.128 [R219+0x5000], desc[UR16][R46.64], P0 ;
IMAD.WIDE.U32 R10, R222, 0x2, R60 ;
LDGSTS.E.BYPASS.128 [R219+0x5800], desc[UR16][R48.64], P0 ;
MOV R33, c[0x0][0x1b0] ;
LDGDEPBAR ;
\end{lstlisting}
\end{minipage}

\newpage
\section{Optimization moves}
\label{appendix. opt-move}


\begin{figure}[htb!]
  \centering
  \includegraphics[width=0.5\linewidth]{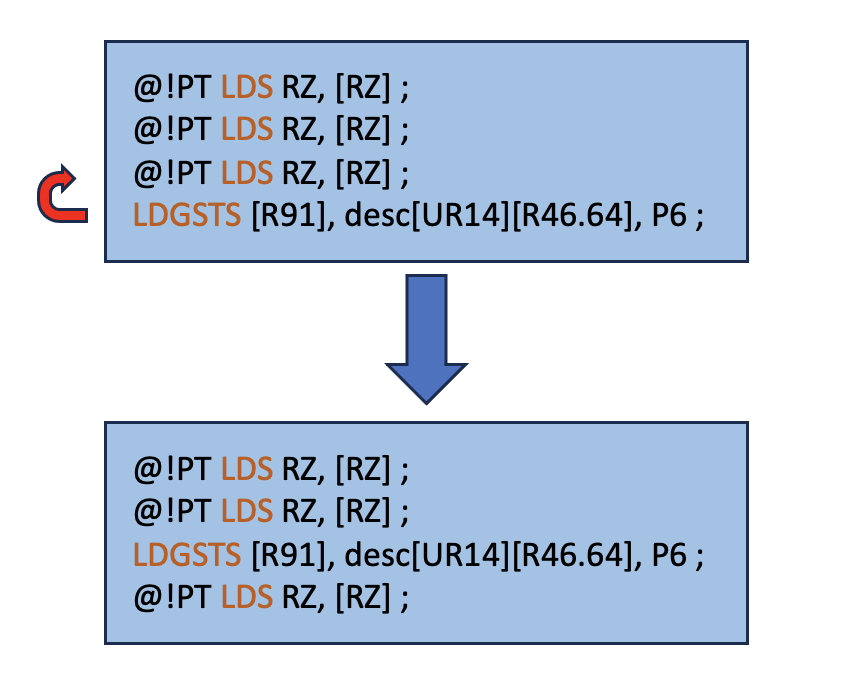}
  \caption{A reordering that schedule the \textbf{LDGSTS} instruction earlier than the predicated \textbf{LDS} instruction. This optimization move is applied to multiple kernels.}
  \label{fig. BMM trace}
\end{figure}

\end{appendices}

\end{document}